\documentstyle[psfig,eqsecnum,aps,preprint,floats]{revtex}
\tightenlines

\def\beq{\begin{equation}}
\def\eeq{\end{equation}}
\def\bea{\begin{eqnarray}}
\def\eea{\end{eqnarray}}
\def\eno#1{Eq.~(\ref{#1})}

\def\Dta{\Delta}

\def\al{\alpha}

\def\lam{\lambda}

\def\baS{{\bar S}}
\def\bH{{\bf H}}

\def\xhat{\bf{\hat x}}

\def\Fe8{Fe$_8$}
\def\Mn12{Mn$_{12}$-ac}

\def\ham{{\cal H}}

\begin{document}

\draft


\title{
Topological Quenching of Spin Tunneling in Mn$_{\bf 12}$-acetate
Molecules}

\author{Chang-Soo Park$^*$ and Anupam Garg}

\address{Department of Physics and Astronomy, Northwestern University,  Evanston, Illinois 60208}

\date{\today}
\maketitle

\begin{abstract}
We investigate the topological quenching of spin tunneling in
Mn$_{12}$-acetate molecules with an applied magnetic field along
the hard axis. The model Hamiltonian describing this system has a
fourth-order term due to the tetragonal anisotroy. We treat this
model using the discrete phase integral formalism in which the
Schr\"{o}dinger equation corresponding to the Hamiltonian becomes
a nine-term recursion relation. We solve this recursion relation
and find that the tunnel splitting is quenched at certain values
of the applied field. We also present a qualitative treatment
based on the instanton approach.
\end{abstract}


\section{INTRODUCTION}
\label{into}

Quantum  tunneling of a spin or spin-like degree of freedom has
been discussed for over a decade now \cite{Gunther}, but
unambiguous evidence for its existence has only come recently
\cite {science} from studies on the magnetic molecule
[Fe$_8$O$_2$(OH)$_{12}$(tacn)$_6$]$^{8+}$ (\Fe8 for short). This
molecule has a total spin $S=10$, is biaxially symmetric, and can
be modeled by the spin Hamiltonian
\begin{equation}
\ham  =  k_1 S_x^2 + k_2 S_y^2 - g\mu_B {\bf S}\cdot\bH,
\label{hamfe8}
\end{equation}
with $k_1 > k_2 > 0$. In the first approximation, it has two
degenerate ground states, approximately given by $S_z = \pm 10$,
which are separated by an energy barrier in the {\it xy}-plane.
The question of interest is to understand how these states are
admixed by quantum tunneling.

Direct numerical diagonalization of \eno{hamfe8} using the
experimentally determined values $k_1 \simeq 0.33$~K, $k_2 \simeq
0.22$~K \cite{schulz,sang,ohm} reveals that the tunnel splitting
$\Dta$ is $\sim 10^{-9}$~K, which is too small to be observed
directly. Wernsdorfer and Sessoli \cite{science} overcome this
difficulty by applying a small amplitude ac magnetic field along
the $z$ direction, which causes the $S_z = \pm 10$ levels to cross
one another. Transitions between these levels are now possible via
the Landau-Zener-St\"uckelberg (LZS) process\cite{LZS}, and the
underlying tunneling matrix element $\Dta$ can be deduced from a
measurement of the incoherent LZS relaxation rate for the total
magnetization. The key experimental fact that supports this
interpretation of the relaxation (which could after all be due to
a classical activation process a priori), is a systematic and
remarkable oscillatory dependence of the inferred splitting $\Dta$
on the strength of the magnetic field $\bH$ when this field is
applied along the hard direction $\xhat$. This phenomon was
predicted some time ago \cite{Garg1} based on an instanton
approach. Briefly, when $\bH
\parallel \xhat$, there are two symmetry related instanton paths that
wind around $\xhat$ in opposite directions, and together form a
closed loop on the (complexified) unit sphere. The actions for the
instantons are complex, and differ by a real valued Berry phase
given by $S$ times the area of the loop, giving rise to
interference \cite{Loss1}.  Now, however, the Berry phase is not
fixed at $2\pi S$, but may be continuously varied by varying $H$.
As a function of $H$, therefore, the tunnel splitting oscillates,
and is completely quenched at values of $H$ where the Berry phase
is an odd integer times $\pi$.

In this paper, we wish to study topological quenching of tunneling
in a second magnetic molecule, Mn$_{12}$-acetate (or \Mn12 for
short), which has also been the subject of several experimental
studies \cite{novak,Fried,Thomas,Hill,Barra}. The reason for our
interest in \Mn12 is that, in contrast to \Fe8, it has tetragonal
symmetry. The spin Hamiltonian of \Mn12 can be written as \beq
{\cal H} = -AS_z^2 - BS_z^4 + C(S_{+}^4 + S_{-}^4 ), \label{mn12}
\eeq where $A \gg B \gg C > 0$. The easy axis $z$ now has four
fold symmetry, the hard axes are $\pm x$ and $\pm y$, and the
medium axes are the lines $y = \pm x$ in the $xy$-plane. Here, the
symmetry of a pair of instanton paths is preserved when a magnetic
field is applied along one of the four hard axes. Thus, the
quenched spin tunneling phenomenon is also anticipated in \Mn12.
Whether it can be observed or not depends on how strong the
environmental decoherence is, and is a question that we shall not
investigate here.

To investigate the topological quenching of the spin tunneling in
\Mn12 molecules we use the discrete phase integral (DPI) (or,
discrete WKB) method\cite{Braun,van}. The DPI method has been
applied to spin tunneling problems in a recent series of works by
one of us (AG) \cite{garg2,garg3,gargfive,gargsymm}. This method
is semiclassical just as the instanton approach is, but it is
easier to use for the study of the splittings of higher pairs of
levels. It is particularly well suited to study tunneling when
{\bf H} is not along the hard axis. As in \Fe8, we anticipate that
$\Dta$ will be quenched at a number of such field values
\cite{garg3,villain}, which correspond to non-trivial diabolical
points\cite{diabolo1,diabolo2} in the magnetic field space. In
this paper, however, we shall not consider such general
orientations of the field as the calculations for ${\bf H}
\parallel \xhat$ are already quite complex.

In the following section we present the DPI formalism for the
present model. Unlike the previously studied model for \Fe8, the
Schr\"{o}dinger equation corresponding to the spin Hamiltonian for
Mn$_{12}$-ac becomes a nine-term recursion relation because of the
fourth-order term. We give a systematic analysis for this
recursion relation. We then calculate the tunnel splittings as a
function of the applied field for the first few energy levels. The
results will be compared with those obtained by numerical
diagonalization of the spin Hamiltoian. In Sec. III we give a
qualitative discussion of spin tunneling based on the instanton
approach. This gives a good physical picture of the quenched spin
tunneling and explains some of the interesting features found in
the DPI results. A summary of the results in Sec. IV concludes the
paper.

\section{DPI CALCULATION OF TUNNEL SPLITTINGS}

We consider the spin Hamiltonian in Eq.~(\ref{mn12})  with
magnetic field applied along the $x$ axis. For convenience we
divide the Hamiltonian by $A$ to work with dimensionless
quantities. With this choice we can write
\begin{equation}
{\cal H} = -S_z^2 - \lambda_1 S_z^4 + \lambda_2 [S_{+}^4 + S_{-}^4
] - \bar{S} h_x S_x,
\end{equation}
where $\lambda_1 = B/A$, $\lambda_2 = C/A$, $h_x = H_x
/\bar{S}H_c$ $(H_c \equiv A/g\mu_B)$, and \beq \bar{S}=S +
\frac{1}{2}. \eeq Here, $\mu_B$ is the Bohr magneton, $g=2$, and
$S$ is the spin. Following Ref.~\onlinecite{Barra}, $A/k_B =
0.556$K, $B/k_B = 1.1 \times 10^{-3}$K, $C/k_B = 3 \times
10^{-5}$K, so that $\lambda_1 = 1.98 \times 10^{-3}$, $\lambda_2 =
5.4 \times 10^{-5}$, and $H_c = 0.414$T. Let $|\bf{\hat n}\rangle
= |\theta, \phi \rangle$ be the spin coherent state with maximal
spin projection along the direction $\bf{\hat n}$, with spherical
coordinates $\theta$ and $\phi$. We introduce the classical energy
\begin{eqnarray}
{\cal H}_c(\theta,\phi) &=& \langle \bf{\hat n}|{\cal H}|\bf{\hat
n}\rangle \nonumber
\\ &=& -S^2 \cos^2 \theta - \lambda_1 S^4 \cos^4 \theta + 2
\lambda_2 S^4 \sin^4 \theta \cos 4\phi - S \bar{S}h_x\sin
\theta\cos\phi .
\end{eqnarray}
When $h_x = 0$, ${\cal H}_c$ has minima at $\theta = 0$, $\theta =
\pi$. As $h_x$ is increased, these minima move toward $\theta =
\pi/2$, $\phi = 0$, lying in the $xz$-plane. At a certain critical
field, $h_{x\rm co}$, these minima will merge with each other,
giving rise to a double zero of $\partial {\cal H}_c (\theta,
\phi=0) /\partial \theta$ at $\theta = \pi/2$. By using this
condition, we can show that \beq h_{x\rm co} = \frac{2S}{\bar{S}}
\left( 1 + 4 \lambda_2 S^2 \right). \eeq With the experimental
numbers given above, $h_{x\rm co} = 1.946$ \cite{note0}.

\subsection{Recursion Relation}

The DPI formalism can be started with Schr\"{o}dinger equation in
the $S_z$ representation. Introducing ${\cal H}|\psi \rangle =
E|\psi \rangle$, $S_z |m\rangle = m|m\rangle$, $\langle m |\psi
\rangle = C_m$, and $\langle m |{\cal H} |m'\rangle = t_{m,m'}$,
the Schr\"{o}dinger equation corresponding to the Hamiltonian
(2.1) can be expressed as
\begin{equation}
\sum_{n=m-4}^{m+4} t_{m,n} C_n = E C_m. \label{recur}
\end{equation}
This is a nine-term recursion relation with diagonal terms
$t_{m,m}$ from $S_z^2$ and $S_z^4$, and off-diagonal terms
$t_{m,m\pm 1}, t_{m,m\pm,4}$ which are from the $S_x$ and
$S_{\pm}^4$ parts, respectively. Since there are no $S_{\pm}^2$ or
$S_{\pm}^3$ terms in the Hamiltonian, we have $t_{m,m\pm 2} =
t_{m,m\pm 3} = 0$.

The recursion relation (\ref{recur}) may be interpreted as the
Schr\"{o}dinger equation of an electron in a one-dimensional tight
binding model. That is, we can consider the diagonal and
off-diagonal terms as the on-site energy and hopping terms,
respectively. Once this analogy is recognized, assuming $t_{m,m\pm
\alpha} (\alpha =0, 1,$ or $4)$ vary slowly with $m$, we can treat
the recursion relation within a continuum quasiclassical
approximation or a phase integral method
\cite{gargfive,Braun,van}. With this approximation we can define
smooth functions
\begin{equation}
t_{\alpha}(m) \simeq \frac{1}{2} (t_{m,m+\alpha} +
t_{m,m-\alpha}), \hspace{0.7cm} \alpha = 0, 1, 4.
\end{equation}
For the present model, $t_{\alpha}$'s are given by
\begin{eqnarray}
t_0 (m) &=& -m^2 (1 + \lambda_1 m^2 ), \nonumber\\ t_1 (m) &=&
-\frac{\bar{S} h_x}{2} \sqrt{{\bar{S}}^2 - m^2},\\ \label{hops}
t_4 (m) &=& \lambda_2 ({\bar{S}}^2 -m^2)^2,\nonumber
\end{eqnarray}
where we have used the approximation $S(S+1) \approx {\bar{S}}^2$.
Introducing the DPI wavefunction within the semiclassical
approximation
\begin{equation}
 C_m \sim \frac{1}{\sqrt{v(m)}} \exp \left[i \int^m q(m')dm'
 \right],
\label{eq:wf}
\end{equation}
we have the Hamilton -Jacobi equation
\begin{equation}
 E = {\cal H}_{sc}(q,m) \equiv t_0(m) + 2t_1 (m) \cos q + 2t_4 (m) \cos
 4q,
\label{eq:HJ}
\end{equation}
and the transport equation
\begin{eqnarray}
 v(m) &=& \frac{\partial {\cal H}_{sc}}{\partial q}\nonumber\\ &=& -2 \sin q(m) [t_1
  (m) + 16 t_4(m) \cos q(m) \cos 2q(m)].
\label{eq:tp}
\end{eqnarray}
In Eqs.~(\ref{eq:wf}) and (\ref{eq:tp}), $q(m)$ is a local,
$m$-dependent Bloch wave vector obtained by solving
Eq.~(\ref{eq:HJ}) for $q$ for any given energy $E$. It is very
useful to have a physical picture of these equations. For a given
value of $m$, Eq.~(\ref{eq:HJ}) gives an energy band $E(q)$ which
defines the classically allowed range of energies. In
Fig.~\ref{fg:eband} we show possible $E$ vs. $q$ curves for our
problem. At lower and upper edges of the band the transport
equation shows that $v(m)$ becomes zero because the slope
$\partial E(q)/\partial q$ is zero. This means the band edges are
related to the classical turning points. These are not the only
turning points, however. Such points are more generally defined by
the condition that the velocity $v(m)$ vanishes. This condition
produces additional loci in $E-m$ space, which we call $critical$
$curves$, along with the $m$-dependent band edges. These curves
are crucial to understanding how the oscillating tunnel splitting,
i.e., the quenching effect, appears.

\subsection{Critical Curves}

>From Eq.~(\ref{eq:tp}) the condition $v(m)=0$ is satisfied when $q
= 0$, or $q = \pi$, or $q = q_*$, where $q_*$ is the solution of
\begin{equation}
32{\,}t_4 (m) \cos^3 q_*(m) - 16{\,}t_4 (m) \cos q_*(m) + t_1 (m)
= 0. \label{eq:ae}
\end{equation}
Substituting these into Eq.~(\ref{eq:HJ}) we obtain the following
energy curves for each of the three $q$'s
\begin{eqnarray}
U_0 (m) &=& t_0 (m) + 2t_1 (m) +2t_4(m),\nonumber\\ U_{\pi}(m) &=&
t_0(m) - 2t_1(m) + 2t_4(m),\label{eq:cc} \\ U_{*}(m) &=& t_0(m) +
2t_1(m)\cos q_*(m) + 2t_4(m) \cos 4q_*(m)\nonumber ,
\end{eqnarray}
where $U_0(m) \equiv E(0,m)$, $U_{\pi}(m) \equiv E(\pi,m)$, and
$U_{*}(m)\equiv E(q_*(m),m)$. Whenever a given energy $E$ crosses
one of these curves a turning point occurs. Various types of
turning points depending on the characteristic of the critical
curves have been analyzed in Ref.~\onlinecite{gargfive}. An
interesting feature of this analysis is the existence of novel
turning points inside the classically forbidden region, which is
crucial for the quenching of spin tunneling. The recursion
relation studied there was based on a spin Hamiltonian which
includes terms up to second order, and there were only three
critical curves to be considered. Here, we expect to have up to
five curves, $U_0(m)$, $U_{\pi}(m)$, and up to three $U_{*}(m)$'s
from the cubic equation (\ref{eq:ae}).

In order to proceed further, it is necessary to analyze the
critical curve structure more closely, in particular, its
dependence on $h_x$. To do this, let us first compare $U_0(m)$
with $U_{\pi}(m)$. From Eq.~(2.7) it can be easily seen that
$U_{\pi}(m) > U_0(m)$ since $t_0(m) < 0$, $t_1(m) < 0$, and $t_4
(m) > 0$ for all $|m| < \bar{S}$. Thus, $U_{\pi}(m)$ can be the
upper band edge. However, in order for this to be so we still need
to prove that $U_{\pi}(m) > U_{*}(m)$. This is not obvious.
Indeed, since the Eq.~(\ref{eq:ae}) is a cubic in $\cos q_*$, it
is possible to have complex solutions. These solutions will yield
a complex $U_{*}(m)$, which is not of interest because the
Hamilton-Jacobi equation $E = U_*(m)$ can not then be satisfied. A
careful consideration of the solutions of Eq.~(\ref{eq:ae}) is
therefore necessary.

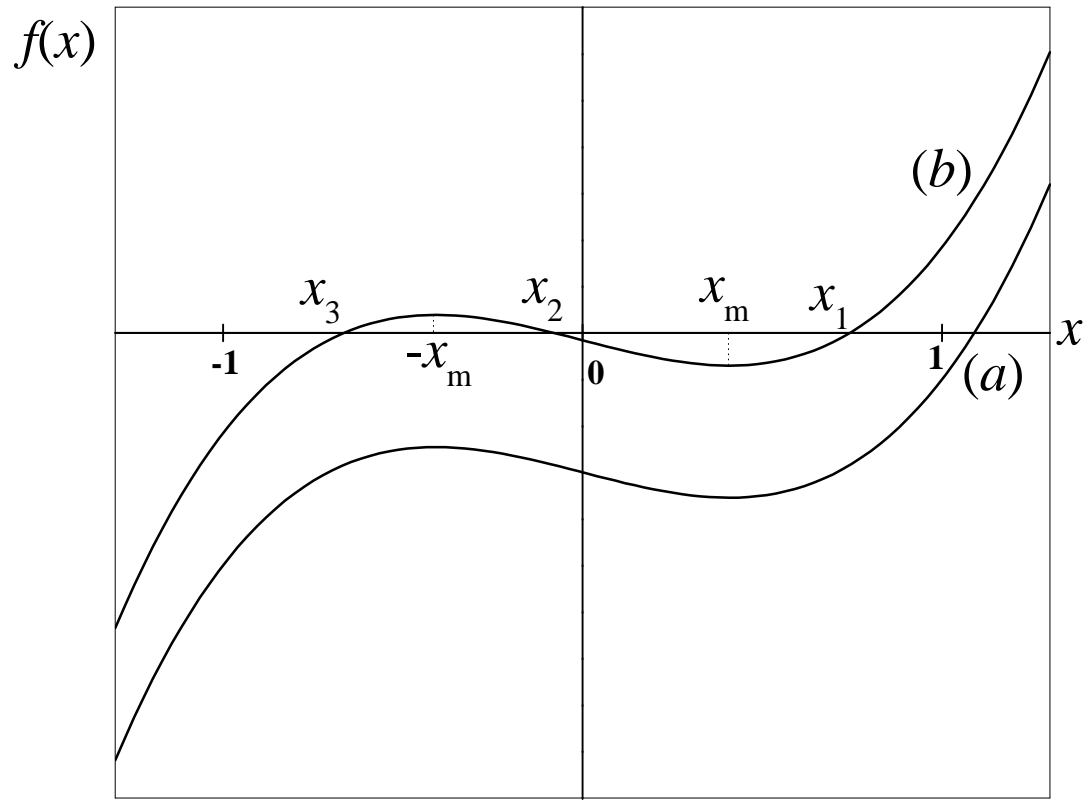
\begin{figure}
\caption{Sketch of the cubic function $f(x)$ for $(a)$ large
$h_x$, or large $|m|$, $(b)$ small $h_x$, or small $|m|$. Note
that, there is one root in $(a)$, but three roots in $(b)$. The
transition from type $(a)$ to type $(b)$ occurs when $f(-x_m) =
f'(-x_m) = 0$ ($x_m = 1/\sqrt{6}$), which gives the curve
$h_{xc}(m)$ in Eq.~(\ref{hxc}). } \label{fg:cubic}
\end{figure}

Defining $x = \cos q_*$, $\mu = m/{\bar S}$, and using Eq.~(2.7)
for the $t_{\al}$'s, we can write Eq.~(\ref{eq:ae}) as
\begin{equation}
f(x) \equiv 2x^3 - x - {h_x \over 32 \lam_2 \baS^2}
                        (1 - \mu^2)^{-3/2} = 0.
   \label{deffx}
\end{equation}
A sketch of the function $f(x)$ is drawn in
Fig.~\ref{fg:cubic}. This sketch incorporates the following easily
verified properties of $f(x)$: (i) $f(0) < 0$, (ii) $f'(0) = -1$,
(iii) $f'(\pm 1) = 5
> 0$, (iv) $f(-1) < 0$, (v) $f(1)$ may be of either sign, (vi)
$f'(\pm 1/\sqrt{6}) = 0$, where $f'(x) \equiv df(x)/dx$. It
follows that a curve of type marked $(a)$, characterized by one
real zero of $f(x)$ arises when $h_x$ is large, or when $|m|$ is
large, and that a curve of type marked $(b)$, characterized by
three real zeros arises when $h_x$ is small, or when $|m|$ is
small. Let us denote the largest zero by $x_1$, and the other two,
when they are real, by $x_2$ and $x_3$ with $x_2 > x_3$. The
corresponding values for $q_*(m)$ and $U_*(m)$ are denoted by
$q_{*i}$ and $U_{*i}(m)$, with $i =1$, 2, or 3. It is obvious that
$ x_1 > 0$, and that $-1 < x_3 < -\frac{1}{\sqrt{6}} < x_2 < 0$.
The first real root yields a positive value for $\cos q_{*1}$, but
since we cannot say if $x_1$ is greater or lesser than 1, $q_{*1}$
may be real or pure imaginary. The other two real roots, when they
exist, always yield real wavevectors $q_{*2}$ and $q_{*3}$.

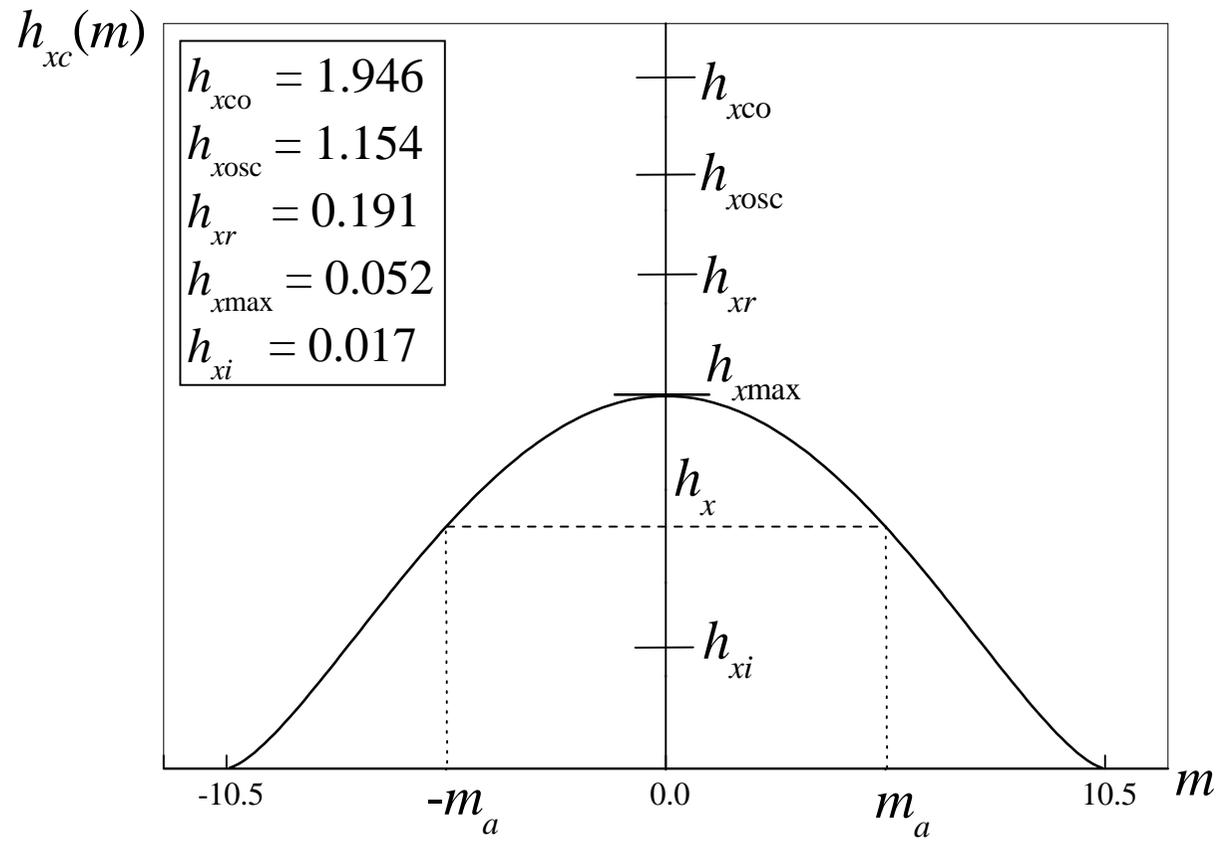
\begin{figure}
\caption{The curve $h_{xc}(m)$ and some physically meaningful
values of $h_x$'s. In the inset we list these values computed with
experimental numbers for $\lambda_1$ and $\lambda_2$ for
Mn$_{12}$-ac. Points at which $h_x$ intersects the curve
$h_{xc}(m)$ are $m = \pm m_a$.}
 \label{fg:hxc(m)}
\end{figure}

The transition from one to three real roots occurs when $f(x)$ has
a double zero, i.e., $f(x)$ and $f'(x)$ both vanish
simultaneously. It is easily shown that this condition is
equivalent to
\begin{eqnarray}
h_{xc}(m) &=& h_{x\rm max} \left(1 -
\frac{m^2}{\baS^2}\right)^{3/2}, \label{hxc}
\\ h_{x\rm max}&=& 32\sqrt{2\over27} \lambda_2{\bar{S}}^2.
\label{hmax}
\end{eqnarray}
The curve $h_{xc}(m)$ and some special values of $h_x$ are
displayed in Fig.~\ref{fg:hxc(m)}. The physical meanings of these
values are listed in Table~\ref{fields}. From the arguments of the
previous paragraph, it follows that we will have three zeros when
$h_x < h_{xc}(m)$, and one zero when $h_x > h_{xc}(m)$. When $h_x
< h_{x\rm max}$, we can also ask for the points $\pm m_a(h_x)$ at
which we change from one to three real roots of $f(x)$. These are
directly given by solving \eno{hxc} for $h_{xc}(m) = h_x$:
\begin{equation} m_a = \baS
\left[1 -
         \left( {h_x \over h_{x\rm max}} \right)^{2/3}
           \right]^{1/2}.
           \label{ma}
\end{equation}

Next, let us investigate whether $U_{*1}(m)$ is inside or outside
the classically allowed energy band. Since $x_1$ moves to larger
positive values as $|m|$ increases (see Fig.~\ref{fg:cubic}), we
see that $U_{*1}$ lies inside the band if $x_1 < 1$, i.e., $|m| <
m_{*}$, where $m_{*}$ is such that $f(1) = 0$. Solving this
equation we get \bea m_{*} &=& \baS \left[1 - \left( {h_x \over
h_{xr}} \right)^{2/3}
           \right]^{1/2},
           \label{mstar} \\
h_{xr} &=& \sqrt{27 \over 2} h_{x\rm max}. \label{hr} \eea
Clearly, $m_a < m_{*}$.

Let us also explore whether the $U_{*i}(m)$'s, when they are real,
are larger or smaller than $U_{\pi}(m)$ or $U_0(m)$. We consider
the following differences:
\begin{eqnarray}
U_{\pi j}(m) &\equiv& U_{\pi}(m) - U_{*j}(m) \nonumber\\ &=&
16t_4(m) \cos q_{*j}(1+\cos q_{*j})^2 (-2+3\cos q_{*j}),
\label{eq:upij}\\ U_{*ij}(m) &\equiv& U_{*i}(m) - U_{*j}(m)
\nonumber \\ &=& -16t_4(m)(\cos^2 q_{*i} - \cos^2 q_{*j} ) \left[
3(\cos^2 q_{*1} + \cos^2 q_{*j} ) - 1 \right], \label{eq:uij}
\\ U_{0i}(m) &\equiv& U_0(m) - U_{*i}(m) \nonumber\\
          &=& 16t_4(m) \cos q_{*i} (1 - \cos q_{*i})^2 (2 + 3\cos
q_{*i}) \label{eq:u0i},
\end{eqnarray}
where $i,j = 1, 2$, or 3 and we have used Eq.~(\ref{eq:ae}) to
eliminate $t_1(m)$ in favor of $t_4(m)$ . From these equations,
and using the facts that $t_4(m) > 0$, plus\cite{note1}
\begin{equation}
-\frac{1}{\sqrt{2}} \leq \cos q_{*3} \leq -\frac{1}{\sqrt{6}} \leq
\cos q_{*2} < 0, \hspace{0.6cm} \frac{1}{\sqrt{2}} \leq \cos
q_{*1}, \label{inequalt}
\end{equation}
we find :

1. When there is only one real root,
\begin{equation}
U_{*1}(m) < U_0(m) < U_{\pi}(m)
\label{eq:uu1}
\end{equation}
for all $|m| < \baS$ and $h_{x\rm max} < h_x < h_{x\rm co}$.

2. When there are three real roots,
\begin{equation}
U_{*1}(m) < U_0(m) < U_{*3}(m) < U_{*2}(m) < U_{\pi}(m)
\label{eq:uu2}
\end{equation}
for $h_{xi} < h_x$,
\begin{equation}
U_{*1}(m) < U_{*3}(m) < U_0(m) < U_{*2}(m) < U_{\pi}(m)
\label{eq:uu3}
\end{equation}
for  $0 < h_x < h_{xi}$, where $h_{xi}$ is determined by
$U_0(m=0,h_{xi}) = U_{*3}(m=0, h_{xi})$, which from
Eqs.(\ref{eq:u0i}) and (\ref{inequalt}), is equivalent to $\cos
q_{*3}(m=0,h_{xi}) = -2/3$ .

We can now list the various types of critical curve patterns that
arise in our problem, and the corresponding ranges of the field
$h_x$. In the following, $U_{-}(m)$ and $U_{+}(m)$ denote the
lower and upper bounds of the energy band, and $U_f(m)$ and
$U_i(m)$ mean the forbidden and internal energies, respectively.

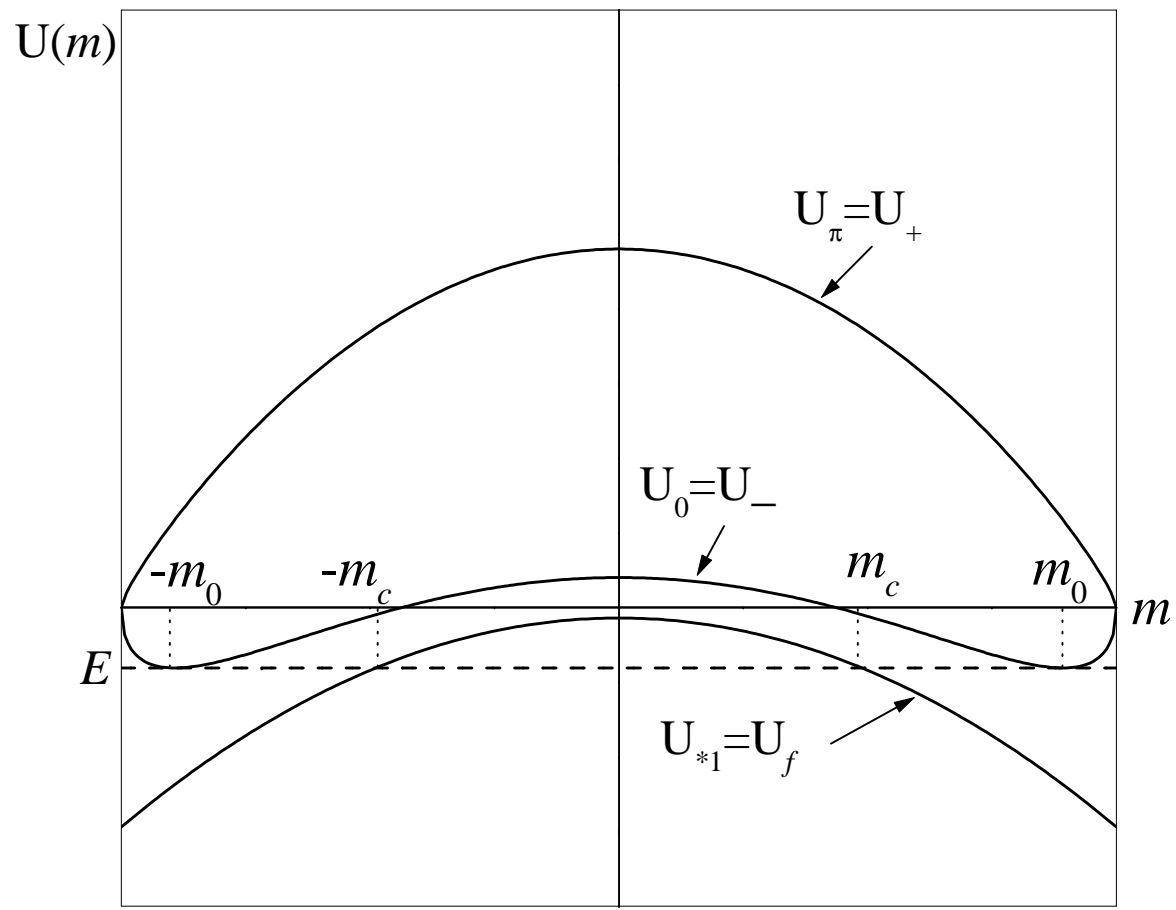
\begin{figure}
\caption{The critical curves for Case~I. At points $\pm m_0$,
$U_0$ has minima, and the points $\pm m_c$ denote the intersection
between $E$ and $U_{*1}$. Note $U_{*1} = U_f$ and $U_0 = U_{-}$
for all $|m| \leq {\bar{S}}$. For a given value of $E$, $q$
becomes complex for $|m| < m_c$ which lies inside the classically
forbidden region. In this region the semiclassical wavefunction
$C_m$ oscillates with decaying or growing envelope.}
\label{fg:cc1}
\end{figure}

Case~I : $h_{xr} < h_x < h_{x\rm co}$.
\newline In this case $U_{*2}$ and $U_{*3}$ are not real for any $m$,
and $q_{*1}(m)$ is imaginary, i.e., $U_{*1}(m)$ is outside the
band for all $|m| \le \baS$. The energy band $E(q)$ is of the type
in Fig.~\ref{fg:eband}a for all $m$, and the critical curves
become
\begin{equation}
U_{*1} = U_f , ~  U_0 = U_{-}, ~ U_{\pi} = U_{+} {\hspace{0.7cm}}
{\rm for}~|m| < \bar{S}, \label{eq:case1}
\end{equation}
which are shown in Fig.~\ref{fg:cc1}.

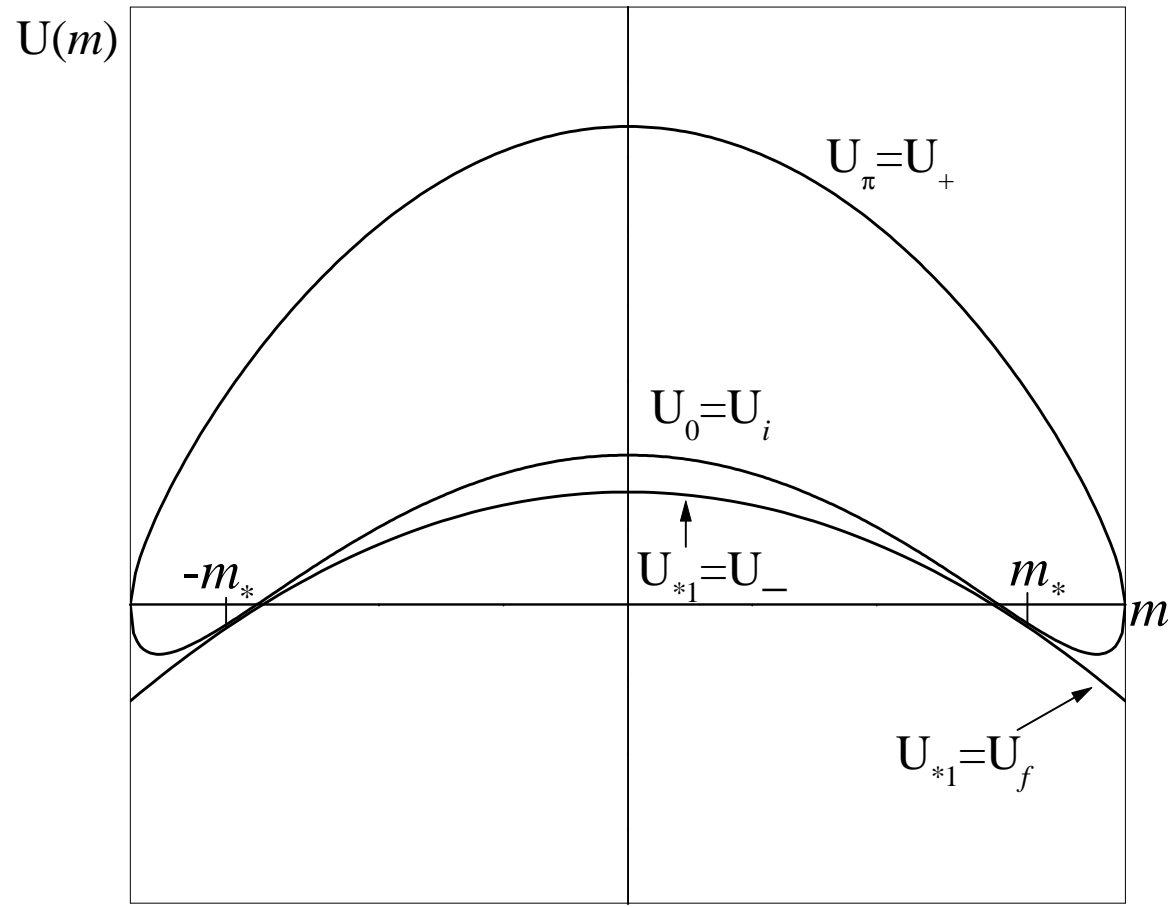
\begin{figure}
\caption{The critical curves for Case~II. $\pm m_*$ are the points
where $U_0(m)=U_{*1}(m)$ [{\it and} $dU_0(m)/dm = dU_{*1}(m)/dm$].
$U_{*1}$ is the lower band edge in the central region $|m| < m_*$
and forbidden in the outer region $|m|
> m_*$.} \label{fg:cc2}
\end{figure}

Case~II : $h_{x\rm max} < h_x < h_{xr}$.
\newline Now, $U_{*2}$ and $U_{*3}$ continue to be complex for
all $m$, but $q_{*1}(m)$ is real in the central region $|m| <
m_{*}$. In this region, the energy band is as in
Fig.~\ref{fg:eband}b, while in the outer region it is of the type
in Fig.~\ref{fg:eband}a. Accordingly, the critical curves have the
structure shown in Fig.~\ref{fg:cc2} and can be written as

\begin{eqnarray}
U_{*1} &=& U_{-},~ U_0 = U_i,~ U_{\pi} = U_{+} \hspace{0.7cm} {\rm
for}~ |m| < m_{*}, \nonumber\\ U_{*1} &=& U_f,~ U_0 = U_{-},~
U_{\pi}=U_{+} \hspace{0.7cm} {\rm for}~ |m| > m_{*}.
\label{eq:case2}
\end{eqnarray}

Case~III : $0 < h_x \leq h_{x\rm max}$.
\newline There are now three $m$ regions. In the outer region,
$|m| > m_{*}$, $U_{*2}$ and $U_{*3}$ are still complex, $U_{*1}$
is outside the band, and $E(q)$ has the shape in
Fig.~\ref{fg:eband}a. In the intermediate range $ m_a < |m| <
m_{*}$, $U_{*2}$ and $U_{*3}$ continue to be complex, but $U_{*1}$
is inside the band, and $E(q)$ has the shape in
Fig.~\ref{fg:eband}b. In the central range, $|m| < m_a$, $U_{*2}$
and $U_{*3}$ become real, and $E(q)$ has the shape shown in
Figs.~\ref{fg:eband}$(c)$ (when $h_{xi} < h_x < h_{x\rm max}$) and
$(d)$ (when $h_x < h_{xi}$). The critical curves can be expressed
as \cite{note2}
\begin{eqnarray}
U_0,U_{*2},U_{*3} = U_i,~U_{*1} &=& U_{-},~ U_{\pi} = U_{+}
\hspace{0.7cm} {\rm for}~|m| < m_a, \nonumber \\ U_0 = U_i,~U_{*1}
&=& U_{-},~U_{\pi} = U_{+} \hspace{0.7cm} {\rm for}~m_a < |m| <
m_*,
\end{eqnarray}
which are illustrated in Figs.~\ref{fg:cc3}a and \ref{fg:cc3}b.

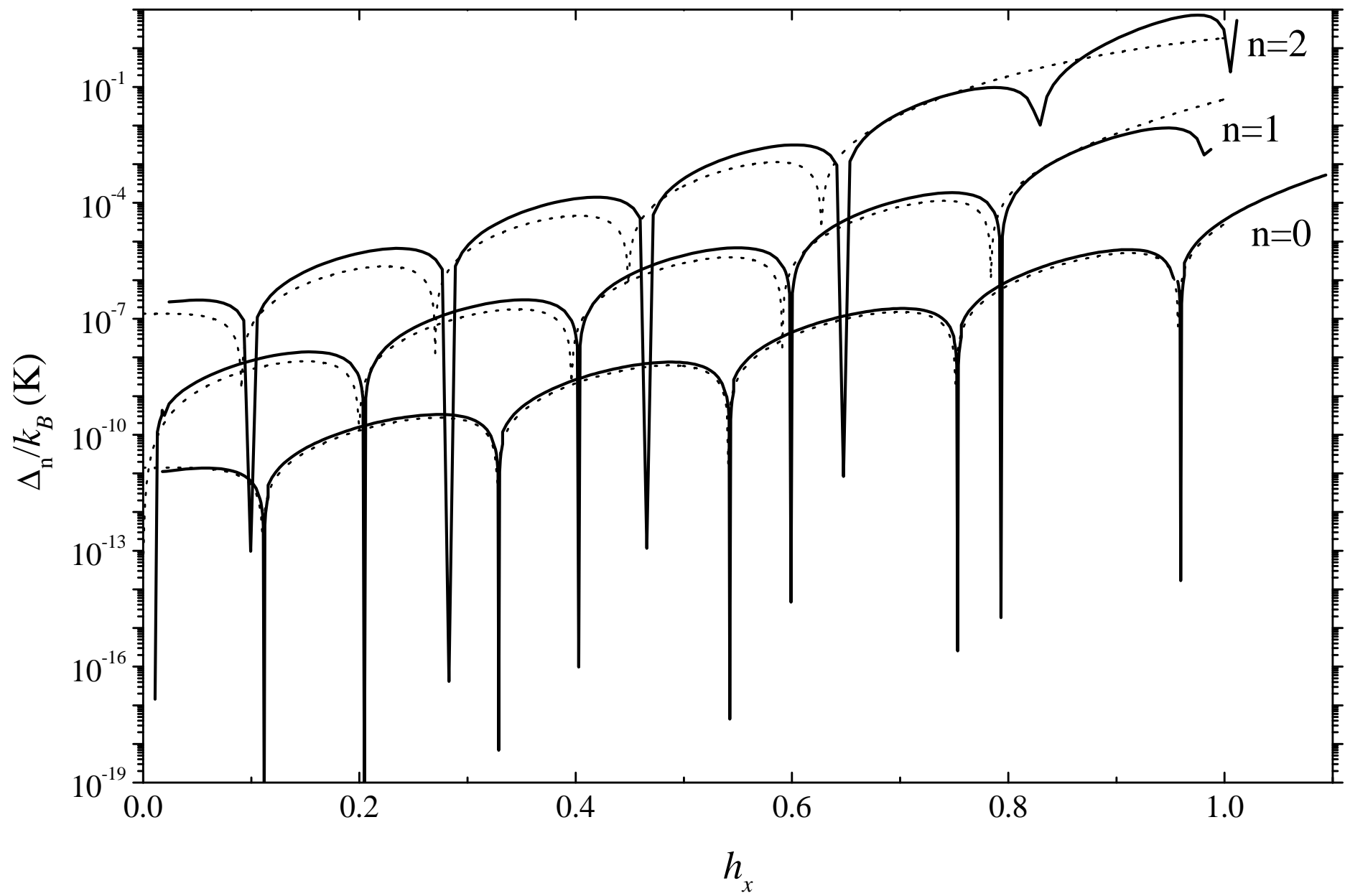
\begin{figure}
\caption{Tunnel splittings $\Delta_n$ between first three pairs of
levels as a function of the field parameter $h_x$. The dotted and
solid curves are obtained from exact numerical diagonalization of
the Hamiltonian and the DPI method, respectively. }
\label{fg:splitting}
\end{figure}

As a matter of fact, we should distinguish two subcases in
Case~III. When $h_x > h_{xi}$, as in Eq.~(\ref{eq:uu2}), the
relevant critical curves are as in Fig.~\ref{fg:cc3}a. When $h_x <
h_{xi}$, as in Eq.~(\ref{eq:uu3}), there is a range of $m$ values
in which $U_{*3} < U_0$ (see Fig.~\ref{fg:cc3}b). For the
experimental parameters relevant to Mn$_{12}$-ac, the field
$h_{xi}$ is rather small, and the points $m_0$, $m_*$, $m_a$, and
$m_i$ are all clustered tightly near $m = \bar{S}$. This means
that for the low lying states, there will be four turning points
very close to one another, and the DPI analysis would have to be
done using a {\it quartic} turning point formula, analogous to the
quadratic turning point formula as discussed by Berry and Mount
\cite{mount}. Since we know the qualitative structure of the
energy spectrum for fields as small as $h_{xi}$, based on the
arguments of Sec.~III, e.g., this exercise is largely academic,
and we have chosen not to perform it. This means that our analysis
is not quite correct at very small fields, and this can be seen in
Fig.~\ref{fg:splitting} especially in the behavior of the
splitting between the first excited pair of levels. As we shall
discuss in Sec.~III, this splitting is rigorously zero at $h_x =
0$, whereas we appear to find a zero at a slightly non-zero value
of $h_x$.

As discussed in Ref.~\onlinecite{gargsymm} the quenching of spin
tunneling occurs when $q(m)$ has a real part as well as an
imaginary part inside the forbidden region. From the viewpoint of
energy curves this happens when there is an energy curve inside
the forbidden region. From the above analysis we can see that only
$U_{*1}(m)$ resides inside the forbidden region. For a given
energy $E$ such that $U_{0\rm min} \leq E < U_{*1\rm max}$, $q$
changes from pure imaginary to complex as $m$ passes from the $|m|
> m_c$ region to the $|m| < m_c$ region, where $m_c$ is the point
where $E$ intersects $U_{*1}(m)$, (for example see
Fig.~\ref{fg:cc1}). When $q$ becomes complex the semiclassical
wavefunction in Eq.~(\ref{eq:wf}) oscillates with exponentially
decaying or growing envelope. The quenching of spin tunneling
arises from this oscillating nature of the wavefuction inside the
forbidden region.

We note here that for the experimental Mn$_{12}$ parameters, the
field $h_{x\rm max}$ is quite small (see the legend in
Fig.~\ref{fg:hxc(m)}), and so in the entire field range for
Case~III, even though there is a forbidden region turning point,
the behavior of the ground state tunnel splitting is qualitatively
similar to that for $h_x = 0$. The behavior of the splitting of
the next two levels is more interesting, and as can be seen from
Fig.~\ref{fg:splitting}, the DPI method does capture it, at least
qualitatively, and perhaps even quantitatively.

\subsection{Tunnel Splittings}

We now calculate the energy splitting due to the spin tunneling
between degenerate states in Mn$_{12}$-acetate. In
Ref.~\onlinecite{gargsymm}, tunnel splittings for five-term
recursion relation have been obtained from Herring's formula. The
final result is, however, quite general so that it can be applied
to a recursion relation which includes more than five terms.
Moreover, as we can notice from the above classifications,
although the present nine-term case has more critical curves the
possible types of the turning points are all included in those
discussed in Ref.~\onlinecite{gargfive}, and no new type of
turning point emerges here. Thus, we can directly apply the
formula for the tunnel splittings obtained in
Ref.~\onlinecite{gargsymm} to the present problem. Since our
calculation is based on this formula we quote the main results
here. The tunnel splitting for $n$th pair of states is given by
\begin{equation}
\Delta_n(h_x) = \frac{1}{n!}\sqrt{\frac{8}{\pi}}\omega_0 F^{n +
\frac{1}{2}} e^{-\Gamma_0} \cos \Lambda_n ,
\label{formula0}
\end{equation}
where
\begin{eqnarray}
\Gamma_0 &=&  \int_{-m_0}^{m_0} \kappa_0(m) dm,\nonumber \\
\Lambda_n &=& \int_{-m_c}^{m_c} \left( \chi_0 + ( n + \frac{1}{2}
) \omega_0 \chi_0^{\prime} \right) dm, \label{formula1}\\ F &=&
2M\omega_0 \left( m_0 - m_c \right)^2 \nonumber \\ &&\times\exp
\left(-2\,Q_1 + \omega_0 \int_{-m_c}^{m_c} \kappa_0^{\prime} dm
\right),\nonumber
\\ Q_1 &=& \int_{-m_0}^{-m_c} \left( \frac{\omega_0
B_0^{\prime}}{\sqrt{B_0^2 - 1}} + \frac{1}{m + m_0} \right)
dm.\nonumber
\end{eqnarray}
Here, $\kappa$ and $\chi$ are the imaginary and real parts of
complex $q$, respectively, and
\begin{eqnarray}
&\kappa_0 = \kappa(m,\epsilon=0); &\kappa_0^{\prime} =\left.
\frac{\partial
\kappa(m,\epsilon)}{\partial\epsilon}\right|_{\epsilon=0},\nonumber
\\ &\chi_0 = \chi(m, \epsilon=0); &\chi_0^{\prime} =\left.
\frac{\partial\chi(m,\epsilon)}{\partial\epsilon}\right|_{\epsilon
= 0},\\ \label{formula2} &B_0 = \cos q(m, \epsilon = 0);
&B_0^{\prime} =\left. \frac{\partial \cos q(m,\epsilon)}{\partial
\epsilon} \right|_{\epsilon = 0},\nonumber
\end{eqnarray}
with $\epsilon \equiv E - U_{-}(-m_0)$. In these equations, $\pm
m_c$ are not quite the turning points of the previous subsection,
in that they are not the point where $U_{*1}(m)$ equals the true
energy $E_n$ of the $n$th pair of levels. Rather, they are the
points where $U_{*1}(m) = U_{-}(\pm m_0)$, which corresponds to
setting $E = U_{-}(m_0)$, i.e., $\epsilon = 0$. The reason is that
the formula (\ref{formula0}) incorporates expansions of various
phase integrals in the energy difference
\begin{equation}
\epsilon_n = E_n - U_{-}(m_0) = (n + \frac{1}{2})\omega_0,
\end{equation}
which is of order $(1/S)$ compared to the energy barrier, as long
as $n \ll S$. This is why $m_c$ is modified, and also why the
primary phase integral for the Gamow factor $\Gamma_0$ runs from
$-m_0$ to $m_0$, the minima of $U_0(m)$, rather than between the
points where $U_0(m) = E_n$. Since all energy curves are a
function of both $m$ and $h_x$, these points still depend on
$h_x$, which in turn makes the $\Delta_n$ depend on $h_x$.

The mass $M$ and frequency $\omega_0$ in Eq.~(\ref{formula1}) are
obtained by approximating $U_{-}(m)$ near its minima by a
parabola, i.e., $U_{-}(m) = E + \frac{1}{2}m\omega_0^2 (m \pm
m_0)^2$. For $m=-m_0$ we find,
\begin{eqnarray}
M &=& - \frac{1}{2t_1(-m_0) + 32t_4(-m_0)},\nonumber \\ \omega_0^2
&=& - 2 [t_1(-m_0) + 16t_4(-m_0)]\left. \frac{\partial^2
U_{-}}{\partial m^2} \right|_{m=-m_0}.
\label{formula3}
\end{eqnarray}

The application of the formulas (\ref{formula0})-(\ref{formula3})
cannot be carried out in closed form all the way, and we must
resort to numerical methods. We explain the principal steps in our
numerical calculation below.

In step 1, we must find $\pm m_0$, and $U_{-}(\pm m_0)$. For our
problem we discover that $U_{-}(m)$ is always given by $U_0(m)$
near the classically allowed minima. The equation for the minima
can be reduced to another cubic,
\begin{equation}
{\bar{S}}^2 h_x^2 = 4({\bar{S}}^2 - y)[2(\lambda_1 - 2\lambda_2)y
+ 1 + 4\lambda_2 {\bar{S}}^2 ]^2,
\label{u0min}
\end{equation}
where $y = m^2$. For th parameters $\lambda_1$ and $\lambda_2$ of
interest to Mn$_{12}$, and $h_x < h_{x\rm co}$, all three roots of
this cubic equation are real, but only one is positive. This root
gives us $m_0$, and substitution of this value into
Eq.~(\ref{eq:cc}) for $U_0(m)$ gives $E$, and
Eqs.~(\ref{formula3}) then give $M$ and $\omega_0$.

Step 2 is to obtain the points $\pm m_c$ given by the roots of the
equation
\begin{equation}
U_{*1}(m) = U_{-}(m_0).
\label{m*}
\end{equation}
As discussed after Eq.~(\ref{formula2}), up to terms of relative
order $(1/S)$, the points $\pm m_c$ are the actual turning points
for the low lying energies. Note that it is $U_{*1}$ which appears
in Eq.~(\ref{m*}) since this is the critical curve that lies in
the classically forbidden region.

To solve Eq.~(\ref{m*}) numerically, we first solve
Eq.~(\ref{eq:ae}) for the function $\cos q_{*1}(m)$, which can be
done in closed form. This solution is then substituted in
Eq.~(\ref{eq:cc}) to obtain $U_{*1}(m)$. The entire procedure can
be implicitly implemented in the numerical routine. The same holds
for $dU_{*1}(m)/dm$. Since $U_{-}(m_0)$ is known from step 1, any
of the standard root-finding methods---Newton-Raphson, bisection,
secant etc.---can be applied to Eq.~(\ref{m*}).

Step 3 is to find $q(m)$, in particular its real and imaginary
parts $\kappa_0(m)$ and $\chi_0(m)$. This is done by solving the
Hamilton-Jacobi equation (\ref{eq:HJ}) with the energy $E$ found
in the first step. The problem amounts to solving a quartic
equation in $\cos q$ and making sure that one has the correct
solution, which can be done easily by making use of the properties
that we have found above. Thus in the region $m_c < |m| < m_0$,
there are two solutions of the form $i\kappa$ (with $\kappa$
real), and two of the form $\pi - i\kappa$. We discard the latter,
and of the former select that one which continuously tends to 0 as
$m \rightarrow \pm m_0$. In the region $|m| < m_c$, the solutions
can be written as $i\kappa \pm \chi$, and as $\pi - (i\kappa \pm
\chi)$, where $\chi \rightarrow 0$ as $m \rightarrow \pm m_c$. We
discard the latter two, and read off $\kappa (\equiv \kappa_0)$,
and $\chi (\equiv \chi_0)$ from the imaginary and real parts of
the first two. Note that both $\kappa_0$ and $\chi_0$ are taken to
be positive.

Step 4 is to find the $\epsilon$ partial derivatives
$\kappa_0^{\prime}$ and $\chi_0^{\prime}$, in effect $\partial
q(\epsilon,m)/\partial\epsilon$. ($B_0^{\prime}$ is directly
obtainable from $\kappa_0^{\prime}$.) We differentiate the
Hamilton-Jacobi equation with respect to $E$:
\begin{equation}
-2[t_1 \sin q + 4{\,}t_4 \sin 4q ] \frac{\partial q}{\partial
\epsilon} = 1.
\end{equation}
Since $q(m)$ is found in step 3, this equation gives $\partial q
/\partial\epsilon$ for any $m$ directly.

We now have all the ingredients needed to evaluate the
one-dimensional integrals $\Gamma_0$, $\Lambda_n$, $F$, and $Q_1$.
This is a straightforward numerical procedure. The only point
worth noting is that the integrand for $Q_1$ is non-singular at $m
= -m_0$, and behaves, in fact, as $(m + m_0 )$.

In Fig.~\ref{fg:splitting} we show the tunnel splittings for first
three pairs of states as a function of the field parameter $h_x$.
For comparison we also plotted the tunnel splittings obtained from
exact diagonalization of the Hamiltonian.

>From the results we observe several interesting features. First,
as anticipated the tunnel splittings are completely suppressed at
certain values of $h_x$. The overall pattern of zeros, their
number, and the dependence of this number on $n$, the pair index,
is understandable on general grounds as we shall discuss. What is
surprising is how regularly spaced these values of $h_x$ are. For
the first pair of splittings, e.g., the intervals between
successive zeros decrease by 2 or 3\% only, and the last interval
is 92\% of the first. For the next pair, $\Delta_2$, the last
interval is 95\% of the first. The mean interval between zeros for
the first three pairs is ${\Delta}H_0 \simeq 0.93$T, ${\Delta}H_1
\simeq 0.85$T, ${\Delta}H_0 \simeq 0.79$T.

The regularity of the zeros means that the phase integral
$\Lambda_n$ decreases almost linearly with $h_x$. (From
Fig.~\ref{fg:splitting}, Gamow factor $\Gamma_0$ also appears to
be quite linear in $h_x$.) While this variation is clearly
expected to be smooth, we have no a priori way to judge how linear
it will be. A similarly strong regularity of quenching intervals
is experimentally discussed in Fe$_8$. The simplest model
Hamiltonian for Fe$_8$ entails only second order terms in the
components of the spin operator, and in this model, the spacing of
zeros {\it is} exactly equal \cite{ersin}, but to describe actual
Fe$_8$, one must add fourth-order terms. These terms change the
spacing significantly, but still seem to preserve its regularity.
It would be interesting to find a physical argument for this
feature, which appears to be somewhat general.

Second, at $h_x = 0$ the tunnel splitting alternates between zero
and nonzero as the level number goes up. This is due to the
fourth-order terms in the spin Hamiltonian. This term causes the
tunneling in the Mn$_{12}$ system, but it also restricts the
transitions to the case that the difference, $\Delta m=|m-m'|$,
between levels $E_m$ and $E_{m'}$ is a multiple of 4. For
tunneling between degenerate states $E_m$ and $E_{-m}$ this
requires the condition $2m=4p$, where $p$ is an integer.
Therefore, there is no tunneling between $+m$ and $-m$ when $m$ is
an odd number.

To give a more detailed argument of this point, we note that when
$H_x = H_z = 0$, because of the $S_{\pm}^4$ terms, the Hamiltonian
can be divided into the following subspaces for $S = 10$.
\begin{eqnarray}
V_1 &=& \{ -10, -6, -2, +2, +6, +10\},\nonumber\\ V_2 &=& \{-9,
-5, -1, +3, +7\},\nonumber\\ V_3 &=& \{-8, -4, 0, +4, +8\},\\ V_4
&=& \{-7, -3, +1, +5, +9\},\nonumber
\end{eqnarray}
where the numbers in brackets give the $m$ quantum numbers. The
subspace $V_1$ contains 6 levels, which form 3 pairs split by
tunneling due to the $CS_{\pm}^4$ terms. The space $V_3$ contains
5 levels, of which $\pm 8$, and $\pm 4$ are split by tunneling,
and $m = 0$ is isolated. There is no degeneracy amongst the 5
levels in space $V_2$, but because of time reversal, this space is
isomorphic to $V_4$, and we therefore conclude that in the full
spectrum of ${\cal H}$, there should be five pairs of {\it
strictly} degenerate levels, corresponding approximately to $m =
\pm (2n+1)$ with integer $n$.

Third, there are five quenching points in the ground and first
excited states tunnelings, and the number of quenching points
decreases as the level number goes up. For the ground state
tunneling the allowed number of quenching points can be explained
qualitatively by thinking in terms of instantons. Since the
instanton approach also gives a good geometrical structure to the
quenching we present it in a separate section.

\section{QUALITATIVE TREAMENT}

In this section we give a qualitative treatment of the spin
tunneling in Mn$_{12}$ using instanton methods. Adding a hard axis
field to Eq.~(\ref{mn12}), the Hamiltonian becomes
\begin{equation}
{\cal H} = -AS_z^2 -BS_z^4 + C(S_{+}^4 + S_{-}^4) - g\mu_B H_x
S_x. \label{haminstan}
\end{equation}
The instanton method is based on spin coherent state path
integrals. In the spin coherent state representation the
anisotropy energy corresponding to the Hamiltonian is given by
\begin{eqnarray}
{\cal H}_c(\alpha, \beta) &=& \langle {\bf{\hat n}}|{\cal H}
|{\bf{\hat n}}\rangle
 \nonumber \\ &=& -AS^2 \sin^2 \alpha \sin^2 \beta -BS^4 \sin^4 \alpha
 \sin^4\beta \nonumber\\ & &+
2CS^4 (\cos^4 \alpha + \sin^4 \alpha \cos^4 \beta - 6 \sin^2
\alpha \cos^2 \alpha \cos^2 \beta) - g\mu_B H_x S \cos \alpha,
\label{aniso}
\end{eqnarray}
where $\alpha$ and $\beta$ are the polar and azimuthal angles with
respect to ${\bf{\hat x}}$, i.e.,
\begin{eqnarray}
\cos\alpha &=& {\bf{\hat n}} \cdot {\bf{\hat x}},\nonumber\\
\sin\alpha \cos \beta &=& {\bf{\hat n}} \cdot {\bf{\hat y}} \\
\sin \alpha \sin \beta &=& {\bf{\hat n}} \cdot {\bf{\hat
z}}.\nonumber
\end{eqnarray}
The energy (\ref{aniso}) exhibits two degenerate minima at
${\bf{\hat n}}_i = (\alpha, \beta)=(\alpha_0, \pi/2)$ and
${\bf{\hat n}}_f = (\alpha, \beta)=(\alpha_0 , -\pi/2)$, where
$\alpha_0 = pi/2$ for $H_x=0$, and decreases smoothly to $0$ as
$H_x$ is increased. The level splitting due to tunneling between
these minima can be obtained from the imaginary time propagator
\begin{eqnarray}
K_{fi} &=& \langle{\bf{\hat n}}_f |\exp[{-\cal H}T] |{\bf{\hat
n}}_i \rangle \nonumber \\&=& \int {\cal D}[{\bf{\hat
n}}]\exp[-S_E [{\bf{\hat n}}(\tau)]],
\end{eqnarray}
 where
\begin{equation}
S_E [{\bf{\hat n}}(\tau)] = -i S {\cal A}[{\bf{\hat n}}(\tau)] +
\int_0^T {\cal{H}}_c(\alpha, \beta) d \tau,
\end{equation}
with
\begin{equation}
{\cal A}[{\bf{\hat n}}(\tau)] =  \int_0^T (1 - \cos
\alpha)\dot{\beta}(\tau) d\tau, \label{eq:geo}
\end{equation}
is the Euclidean action and complex in general. Here, the boundary
conditions are ${\bf{\hat n}}(0) = {\bf{\hat n}}_i$, ${\bf{\hat
n}}(T) = {\bf{\hat n}}_f$. Geometrically, for a closed path, the
integral in Eq.~(\ref{eq:geo}) can be interpreted as a surface
area on the unit sphere enclosed by this path, which can be
verified by Stokes theorem. In the large spin limit the path
integral can be approximated by the sum of all contributions from
paths that minimize the action, that is, the instanton paths. The
instantons for the present model are not simple because of the
fourth-order terms. However, we can construct a qualitative
argument to find the quenching effect without performing explicit
calculations . Since the Euclidean action $S_E$ has both real and
imaginary parts we can generally express the ground state tunnel
splitting as
\begin{equation}
\Delta = \sum_j D_j e^{-S_{Rj}} e^{iS_{Ij}}, \label{eq:delta}
\end{equation}
where $j$ labels the various instantons, $S_{Rj}$, $S_{Ij}$ are
the real and imaginary parts of the instanton action,
respectively, and $D_j$ are prefactors. With these ingredients we
now discuss how the quenching appears in the present model.

Let us first consider the case when $H_x=0$. Since the energy has
four-fold symmetry, an argument of von Delft and Henley can be
applied \cite{Loss1}. If ${\bf{\hat n}}(\tau)$ is an instanton
path, so is ${\cal R}_{\bf{\hat z}}(\pi/2)\bf{\hat n}(\tau)$,
where ${\cal R}_{\bf{\hat z}}(\pi/2)$ is a rotation through
$\pi/2$ about $\bf{\hat z}$. Keeping in mind that $\bf{\hat
n}(\tau)$ is complex, when we project onto the real unit sphere,
there are four saddle point paths passing through each of the four
medium directions. Because of symmetry, each has the same real
contribution to the action integral $S_R$. However, since their
azimuths about the easy axis are different, the imaginary part of
the action, i.e., the phase $S_I$, will not be the same. From the
geometrical meaning of the integral in Eq.~(\ref{eq:geo}), the
phase difference between two instanton paths equals $S$ times the
surface area on the unit sphere enclosed by these instanton paths.

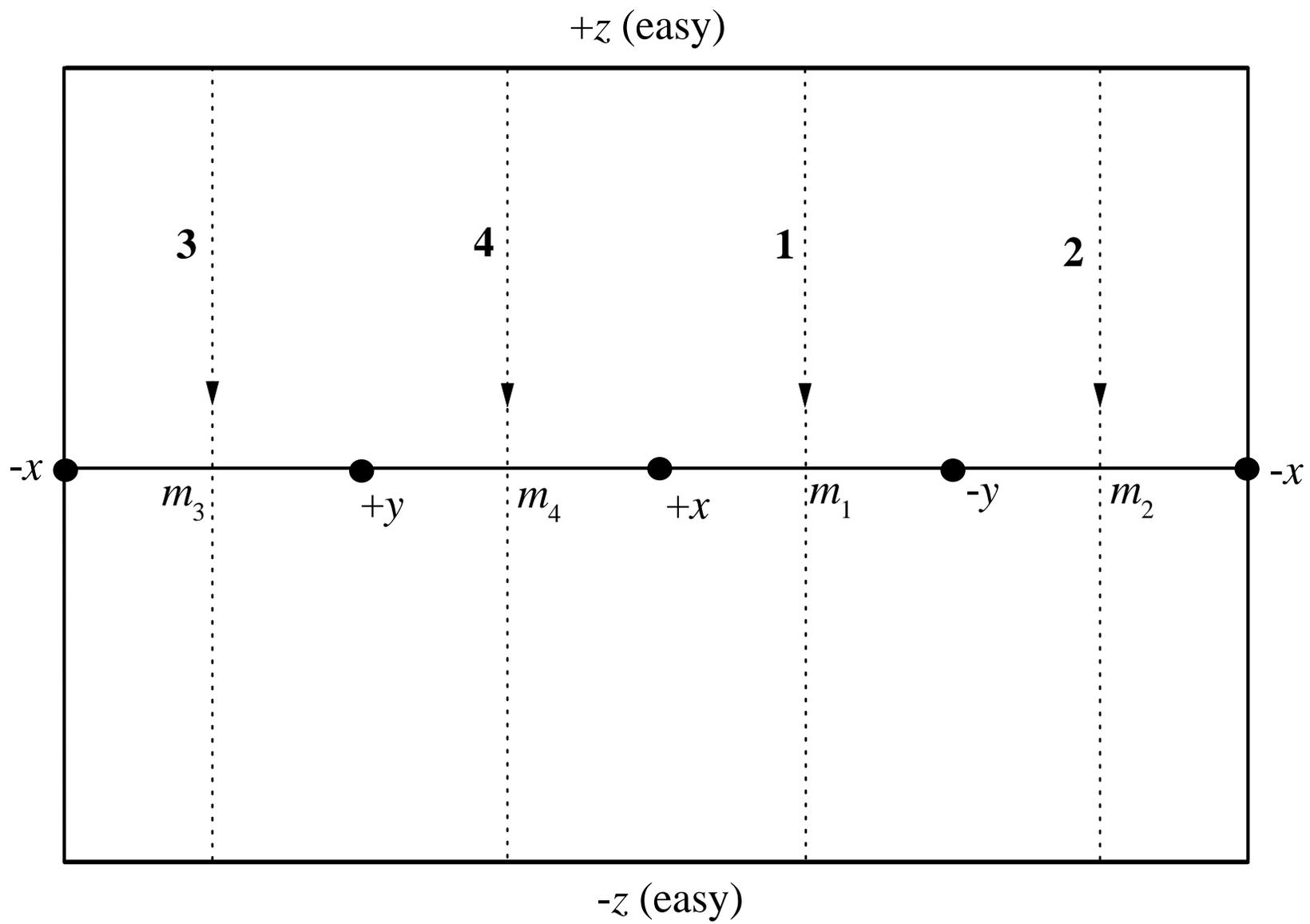
\begin{figure}
\caption{Two dimensional picture of instanton paths when $H_x =
0$. The points $\pm x$, $\pm y$ are the hard axes, and $m_i$'s
represent the medium axes. Dotted lines denote the real
projections of the instanton paths.} \label{fg:phase1}
\end{figure}

To visualize the interference effect we map the two-sphere onto a
plane, as in an ordinary atlas (see Fig.~\ref{fg:phase1}). The
hard axes are mapped onto four equally separated points lying on
the equator, and the points exactly halfway between these
correspond to the medium axes. Thus, the real projections of the
instanton paths can be drawn as curves which start from $+z$, pass
through the medium points, and end at $-z$. The area enclosed by
two adjacent instanton paths equals $\pi$, since the sphere is
equally divided into four paths by the instantons. Thus, the phase
difference between adjacent paths becomes $S\pi$. Choosing the
phase of path 1 as the base, we can perform the summation in
Eq.~(\ref{eq:delta}). Recalling that by symmetry the contribution
from real parts of the instanton actions are all same, as are the
prefactors $D_j$, we have
\begin{eqnarray}
\Delta &=& D e^{-S_R} e^{-iS_I} (1 + e^{-i\pi S} + e^{-2i\pi S} +
e^{-3i\pi S}) \nonumber \\ &=&4De^{-S_R} e^{i\gamma} \cos(\pi S)
\cos\left(\frac{\pi S }{2}\right),
\end{eqnarray}
where $\gamma$ is an irrelevant phase. This result gives us two
quenching conditions. From the factor $\cos \pi S$, we obtain the
quenching of spin tunneling for half-integer $S$, which is just
the Kramers degeneracy effect. The second cosine implies that the
ground state spin tunneling is quenched for odd integer spins,
i.e., $S =$ 1,3,5, etc, and so $\Delta$ is non-zero only for $S =
2p$, where $p$ is an integer\cite{note3}.

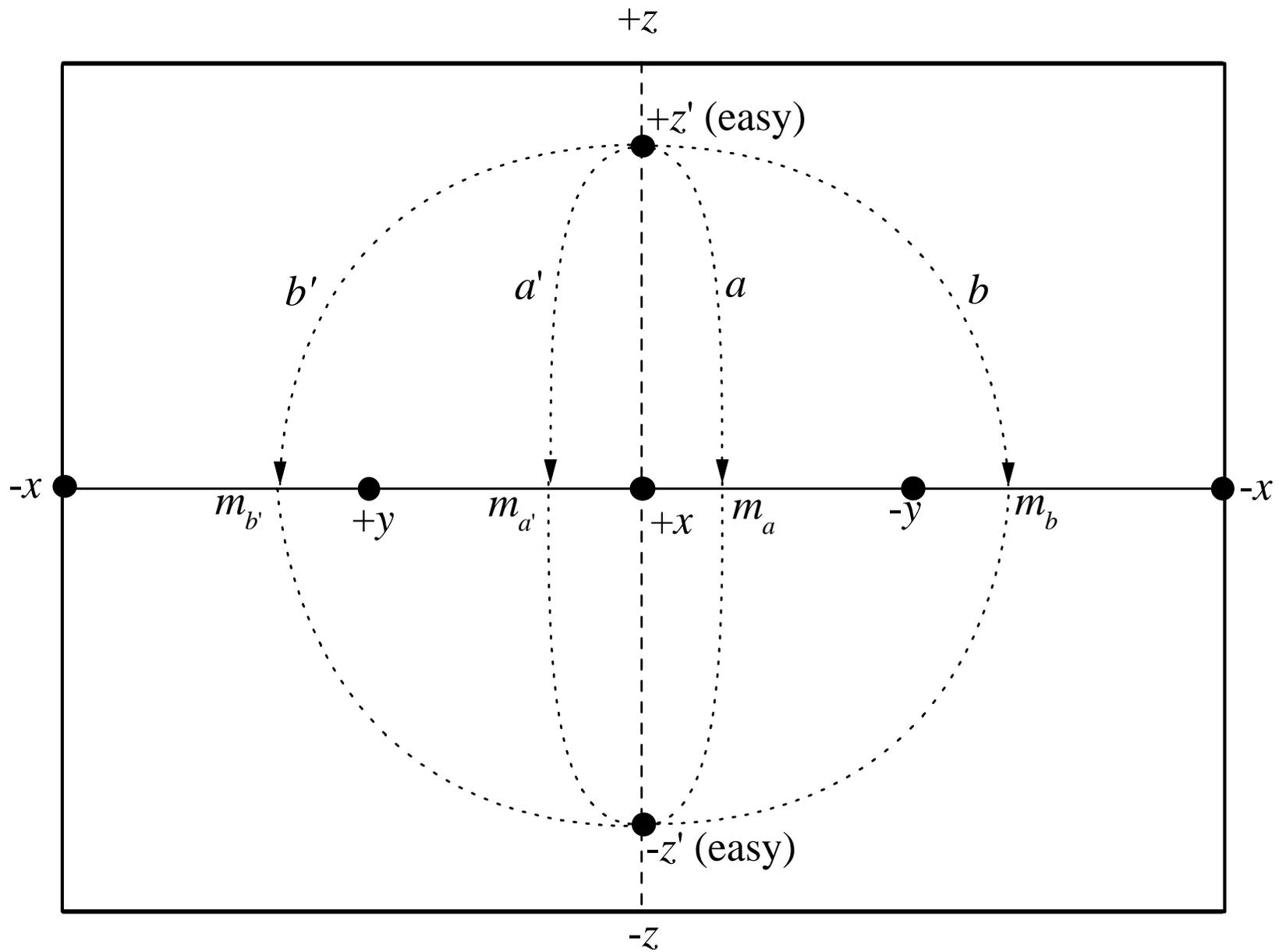
\begin{figure}
\caption{Two dimensional picture of instanton paths when $H_x \neq
0 $. The points $\pm z'$ represent the new easy axes. The
instanton paths are again denoted by dotted lines. Note that the
areas enclosed by each pair of instanton paths are shrunk due to
the field. } \label{fg:phase2}
\end{figure}

We now consider the case with $H_x \neq 0$. Since the field is
assumed to be applied along the $+x$ axis, both easy and all four
medium axes move close to $+x$ axis. Thus, the two dimensional
picture becomes one as shown in Fig.~\ref{fg:phase2}. The
four-fold symmetry is now broken, but there are two pairs of
instanton paths surrounding the $+x$ axis: $(a, a')$ and $(b,
b')$. The real parts of the instanton actions in a pair are same,
but different between the pairs. The phase differences in each
pair are the areas enclosed by each pair of instanton paths (the
small and large oval regions in Fig.~\ref{fg:phase2}) and are
dependent on the field $H_x$. If we choose the straight line
joining $+z'$ to $-z'$ as a reference, $S_{Ia'} = -S_{Ia}$,
$S_{Ib'} = -S_{Ib}$, so that the summation in Eq.~(\ref{eq:delta})
can be performed as
\begin{eqnarray}
\Delta &=& D_a e^{-S_{Ra}}\left[e^{-iS_{Ia}} + e^{-iS_{Ia'}}
\right] + D_b e^{-S_{Rb}} \left[e^{-iS_{Ib}} +
e^{-iS_{Ib'}}\right] \nonumber
\\ &=& 2D_a e^{-S_{Ra}} \cos \frac{SA_a(H_x)}{2}  +
2D_b e^{-S_{Rb}} \cos \frac{SA_b(H_x)}{2}, \label{eq:delta2}
\end{eqnarray}
where $S_{Ra}$, $S_{Rb}$ are the real parts of the instanton
actions in each pair, and $A_a(H_x)$, $A_b(H_x)$ are the areas
enclosed by the pairs $(a, a')$ and $(b, b')$, respectively. For
$H_x > 0$ the saddle points through which the paths $(a, a')$ pass
are lower than those for $(b, b')$, which means that $S_{Ra} <
S_{Rb}$. The main contribution to $\Delta$ in
Eq.~(\ref{eq:delta2}) then comes from the first term, and we can
neglect the second term. The quenching of the ground state tunnel
splitting thus arises when $A_a(H_x) \simeq (2n+1)\pi / 2S$, where
$n$ is a non-negative integer. To see how many quenching points
are allowed we note that $A_a (H_x) < A_a(0)$, where $A_a(0) =
\pi$ (the area enclosed by the two paths 1 and 4 in
Fig.~\ref{fg:phase1}). From this condition we find $n < (S-1)/2$.
For $S=10$ there are thus five values of $H_x$ at which the
quenching appears.

The instanton argument provides another way of seeing that the
region of very small (but non-zero) $H_x$ is special. Exactly at
$H_x=0$, four instantons are important, but for large $H_x$ only
two are important. There must therefore be a regime of small $H_x$
where we make a smooth transition between these two behaviors. The
width of this regime can be quite small since $S_{Rb} > S_{Ra}$ as
soon as $H_x \neq 0$, and these actions appear in the exponents in
Eq.~(\ref{eq:delta2}), so that the difference $(S_{Rb} - S_{Ra})$
is amplified. We thus have another way of seeing why the formula
(\ref{formula0}) fails near $h_x = 0$. It contains only cosine
factor, and is effectively ignoring the second term in
Eq.~(\ref{eq:delta2}).

\section{SUMMARY}
\label{sum}

In this paper, we have used the DPI method to study tunneling in
Mn$_{12}$, especially its behavior with a hard-axis field, which
is expected to show oscillation as in Fe$_8$. The recursion
relation now has nine terms, complicating the analysis. There may
be up to five critical curves, which leads to many more turning
points. The DPI method still works, however, even though the phase
integrals and integrands must be evaluated numerically. But, the
numerical procedures required are simple, and involve only root
finding and integration in one variable. Except for some special
narrow field regions, where two or more turning points merge, the
DPI analysis based on linear turning point formulas is extremely
good, and agrees with exact numerical results quantitatively.

\newpage

\begin{figure}
\caption{The different types of energy bands possible for
recursion relation (\ref{recur}) with wavefunction (\ref{eq:wf}).
Note $U_{\pi}(m) = U_+$ in all cases. $(a)$ When $h_{xr} < h_x <
h_{x\rm co}$ with $|m| < {\bar{S}}$, and when $h_{xc}(m) < h_x <
h_{xr}$ with $|m| > m_*$. In this case, $U_0(m) = U_{-}$, and
$U_{*1}(m)$ does not appear since $q_{*1}$ is imaginary. $(b)$
When $h_{xc}(m) < h_x < h_{xr}$ with $m_a < |m| < m_*$. Here,
$U_{*1}(m) = U_{-}$, $U_0(m) = U_i$. $(c)$ When $h_{xi} < h_x <
h_{x\rm max}$ with $|m| < m_a$, and when $0 < h_x < h_{xi}$ with
$m_i < |m| < m_a$. $(d)$ when $0 < h_x < h_{xi}$ with $|m| < m_i$.
Note that, in both $(c)$ and $(d)$, $U_{*1}(m) = U_{-}$, and
$U_0(m)$, $U_{*2}(m)$, and $U_{*3}(m)$ are inside the band and
thus all denoted $U_i$.}\label{fg:eband}
\end{figure}

\begin{figure}
\caption{The critical curves for Case~III. $(a)$ When $h_{xi} <
h_x < h_{x\rm max}$ and $(b)$ when $0 < h_x < h_{xi}$. There are
five critical curves. Note, however, that $U_{*2}$ and $U_{*3}$
appear only in the region $|m| < m_a$ because they are complex
outside this region.}\label{fg:cc3}
\end{figure}

\newpage

\begin{table}[ht]
\caption{Physical meanings of the special $h_x$'s} \vspace{0.2cm}
\begin{center}
\begin{tabular}{l|l}
$h_{x\rm co}$   &Coercive field above which no tunneling exists.\\
\hline $h_{x\rm osc}$ &The value below which the wavefunction can
have \\& an oscillating part inside the forbidden region.\\ \hline
$h_{xr}$ &The value above which $q_{*1}$ becomes  real.\\ \hline
$h_{x\rm max}$ &The maximum value of the curve $h_{xc}(m)$. \\
\hline $h_{xi}$ &The value at which $U_{*3}(0)$ intersects
$U_0(0)$.
\end{tabular}
\end{center}
\label{fields}
\end{table}


\begin{thebibliography}{10}
\bibitem{note0}
This field value, along with some other important ones, is listed
in Table I.
\bibitem{note1}
The lower bounds for $x_1$ and $x_3$ are found by noting that they
arise at $h_x = 0$, since for any fixed $m$, the curve for $f(x)$
shifts upward as $h_x$ is reduced. The lower bound for $x_2$ (and
upper bound for $x_3$) follows from the fact that $f'(x) = 0$ at
$x = -1/\sqrt{6}$.
\bibitem{note2}
By writing that $U_0$ and $U_{*2,3} = U_i$, we do not mean that
these three critical energies are equal, but that they lie {\it
inside} the classically allowed region.
\bibitem{note3}
This result also stems from the fourth-order term since it only
allows transitions between energy levels with $\Delta m = 4p$,
where $p$ is an integer. For spin $S$ the existence of the ground
state tunneling requires $2S = 4p$, as before.
\end{thebibliography}
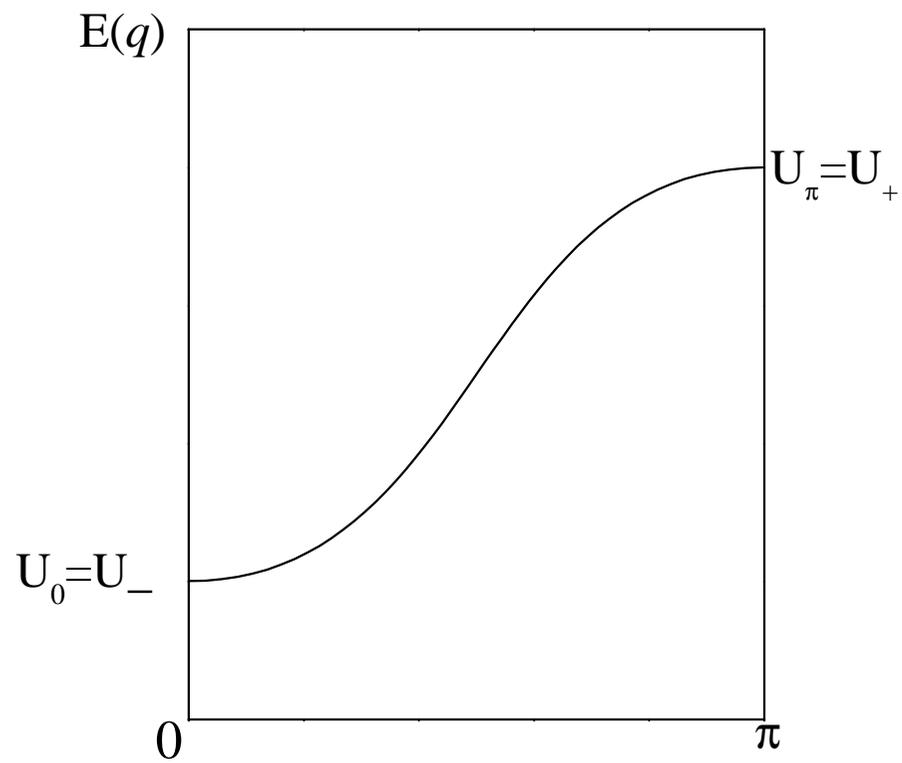
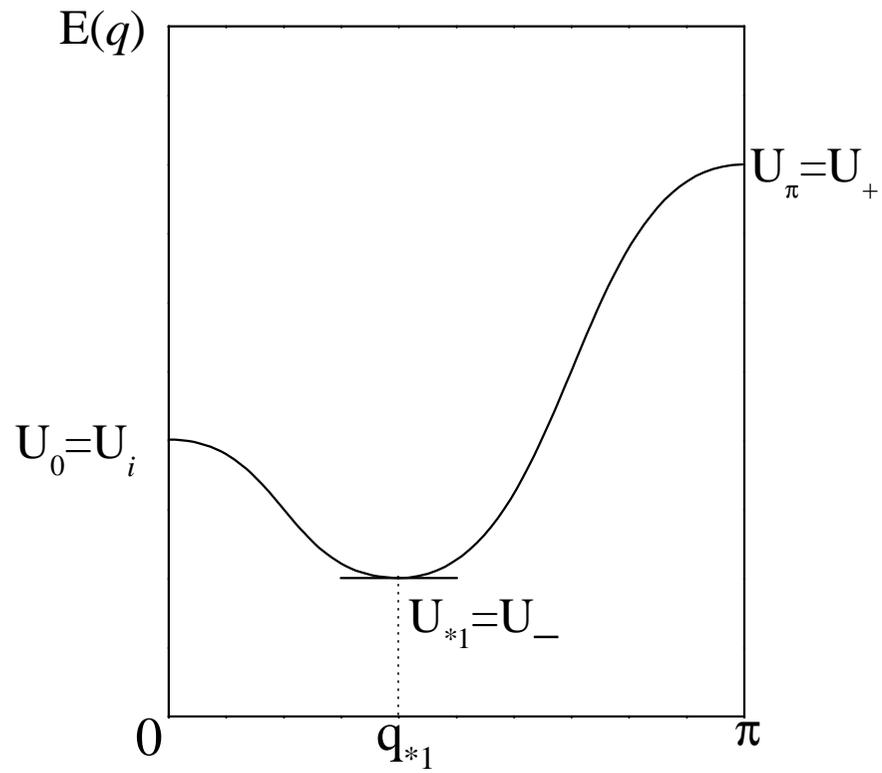
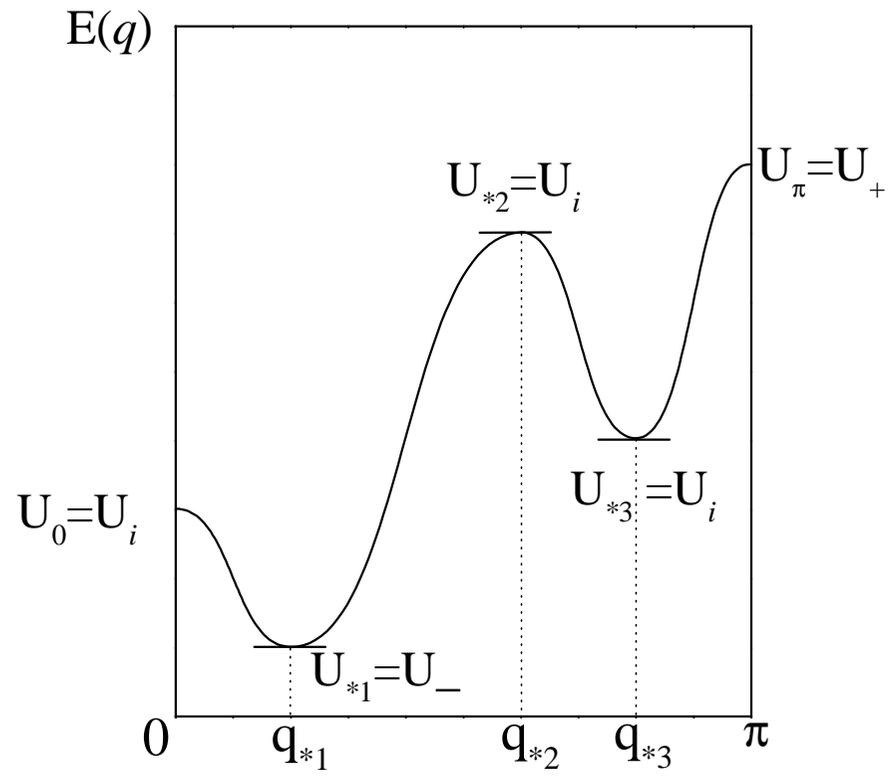
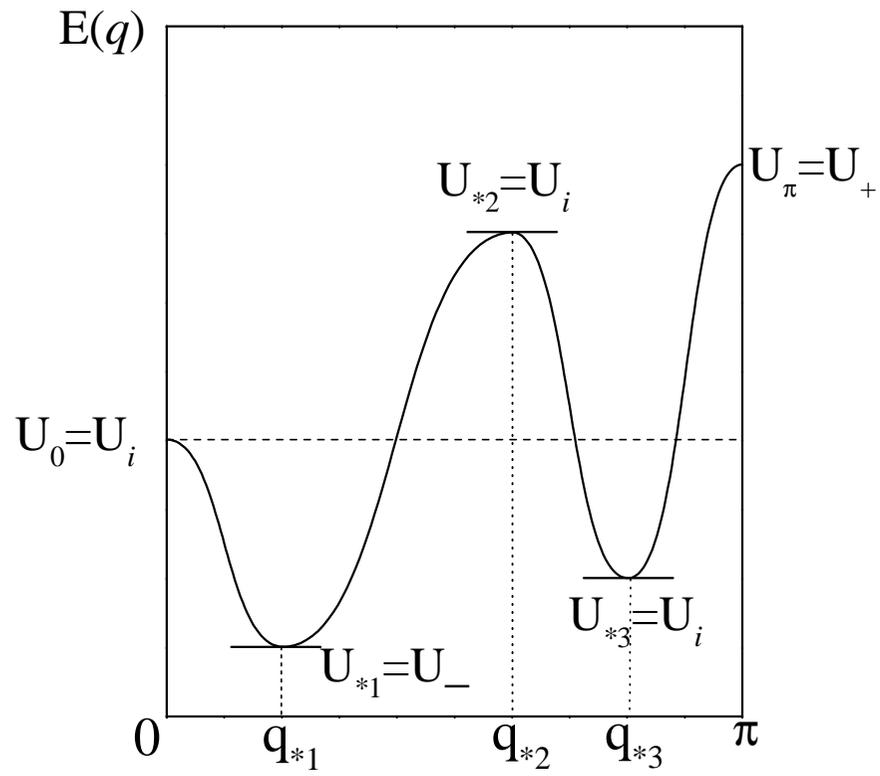
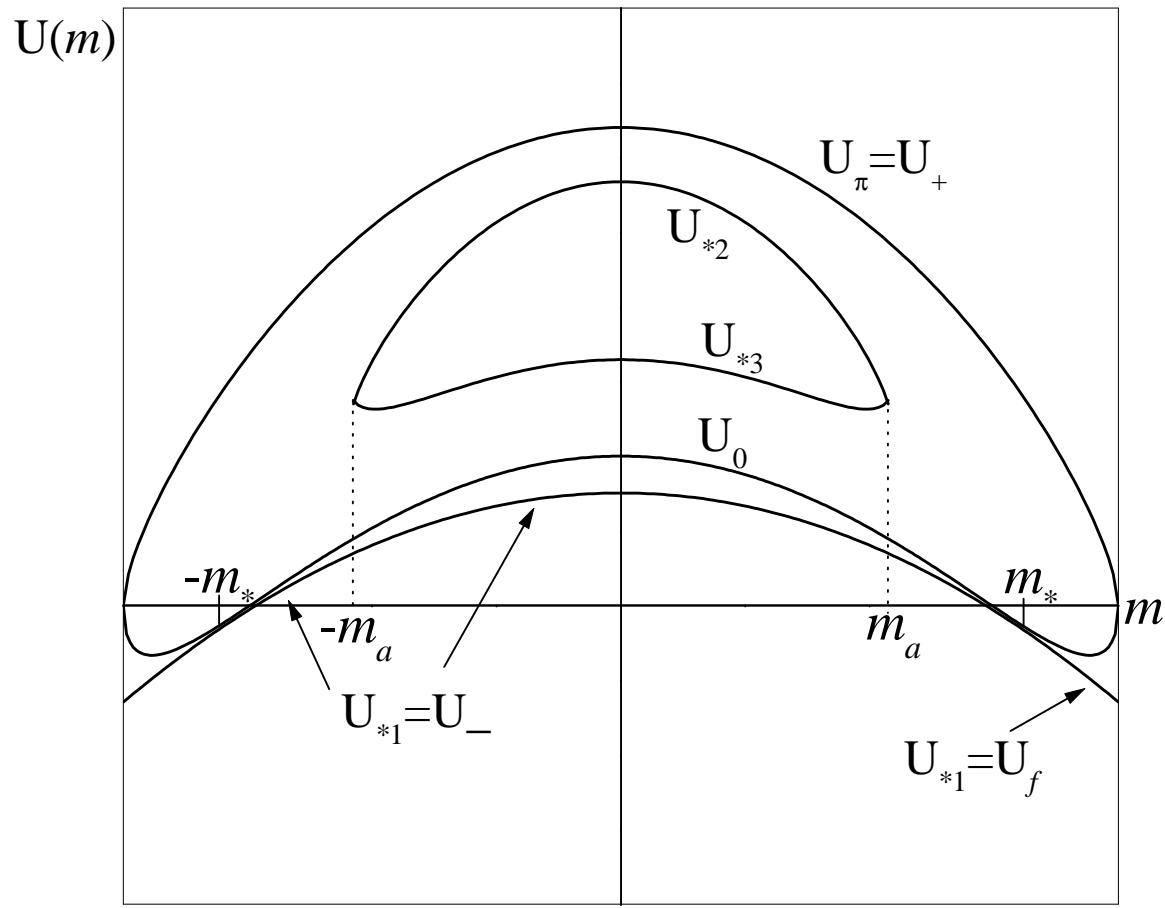
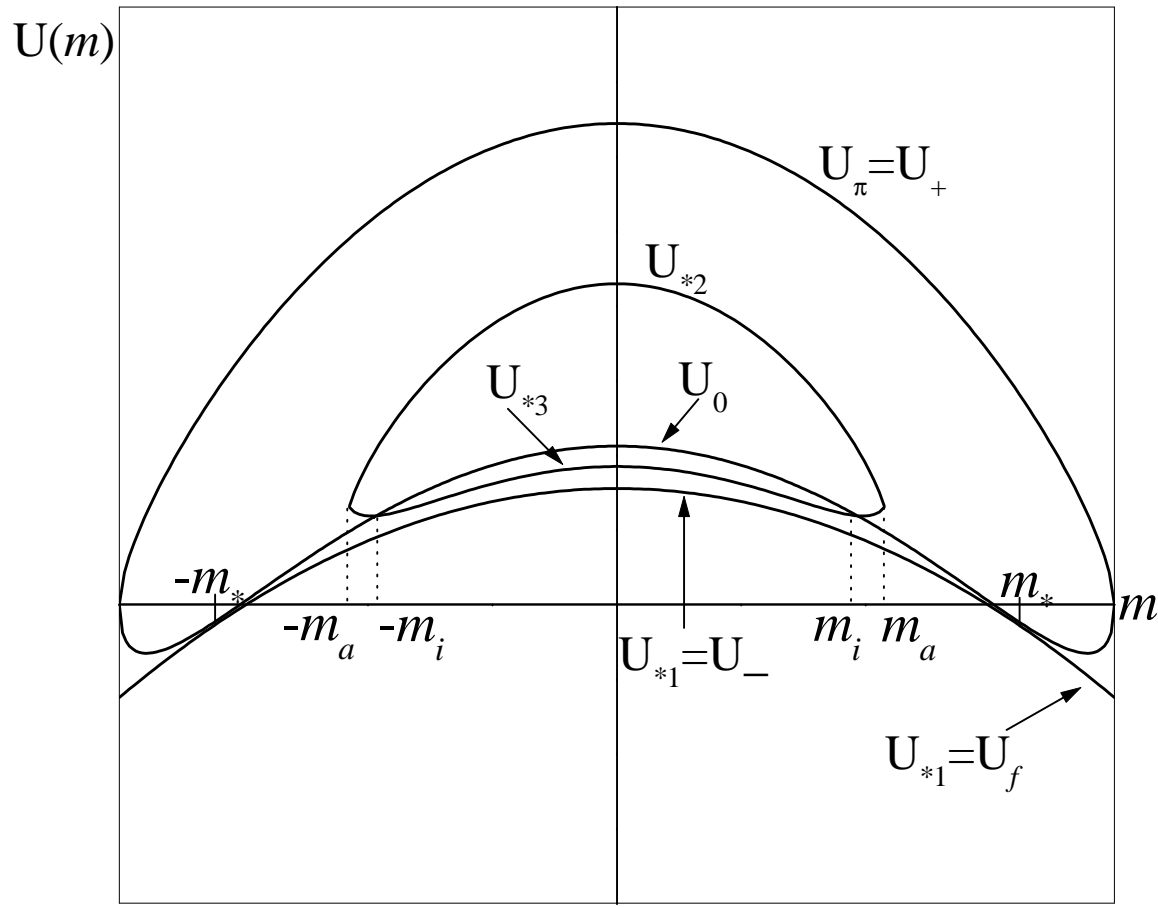
\end{document}